\makeatletter
\declare@file@substitution{revtex4-1.cls}{revtex4-2.cls}
\makeatother
\documentclass[twocolumn,trackchanges]{aastex7}
\hypersetup{linkcolor=red,citecolor=blue,filecolor=cyan,urlcolor=red}

\shorttitle{Nonlinearity of 3 minute Slow Magnetoacoustic Waves}
\shortauthors{Y. Sanjay, S. Krishna Prasad, R. Sych, P.S. Rawat}

\graphicspath{{./}{}}
\begin{document}
\title{Nonlinearity of 3 minute Slow Magnetoacoustic Waves in the Sunspot Umbral Atmosphere}

\author[orcid=0009-0005-2678-5082,gname='Sanjay']{Y. Sanjay}
\affiliation{Department of Physics, School of Advanced Engineering\\
University of Petroleum and Energy Studies, Dehradun-248007, Uttarakhand, India}
\email[show]{ysanjay632@gmail.com}  

\author[orcid=0000-0002-0735-4501,gname='Krishna']{S. Krishna Prasad}
\affiliation{Aryabhatta Research Institute of Observational Sciences (ARIES) \\
Manora Peak, Nainital-263001, Uttarakhand, India}
\email[show]{krishna.prasad@aries.res.in}  

\author[orcid=0000-0003-4693-0660,gname='Robert']{Robert Sych}
\affiliation{Institute of Solar-Terrestrial Physics SB RAS, Irkutsk 664033, Russia}
\email[]{sych@iszf.irk.ru} 

\author[orcid=0000-0003-3485-9430]{P.S. Rawat}
\affiliation{Department of Physics, School of Advanced Engineering\\
University of Petroleum and Energy Studies, Dehradun-248007, Uttarakhand, India}
\email[]{psrawat@ddn.upes.ac.in}

\begin{abstract}
Slow magnetoacoustic waves with a 3 minute period are upward-propagating waves traveling through the density-stratified umbral atmosphere. The decreasing density causes their amplitude to increase, developing into nonlinear waves through steepening and eventually forming shocks. To investigate the vertical evolution of this wave nonlinearity, we utilized multi-wavelength data from the Atmospheric Imaging Assembly (AIA) onboard the Solar Dynamics Observatory (SDO), covering from the photosphere to the lower corona across 20 active regions. The steepening of the wave profile leads to the generation of higher harmonics. We quantify this using a nonlinearity index (NI), defined as the ratio of the amplitude of 2nd harmonic to the fundamental obtained using wavelet analysis. We find a characteristic pattern: nonlinearity increases from the photosphere through the lower chromosphere, peaking near the AIA 1700 \r{A} formation height, and decreases at higher altitudes, notably in the AIA 304 \r{A} channel. This trend indicates progressive wave steepening and subsequent energy dissipation before reaching the formation of AIA 304 \r{A}, consistent with shock formation in the lower atmosphere. An additional rise in NI is observed at the AIA 131 \r{A} channel, followed by a decline in AIA 171 \r{A}, suggesting a 2nd phase of wave nonlinearity evolution in the lower corona. Based on the NI profile and the formation heights of these channels, we conjecture that nonlinear wave processes are most prominent between the AIA 1700 \r{A} and AIA 304 \r{A} formation layers and again between AIA 131 \r{A} and AIA 171 \r{A}.

\end{abstract}


\section{Introduction} \label{sec:intro}

The slow magnetoacoustic waves (MAWs) are regularly observed in the sunspot atmosphere. These waves are thought to be driven by external $p$-mode absorption
\citep[e.g.,][]{2015ApJ...812L..15K}  and magnetoconvection \citep[e.g.,][]{2017ApJ...836...18C}. The dominant oscillatory periods in the sunspots range from 5 minutes in the photosphere to 3 minutes in the chromosphere and the corona \citep[e.g., see review][]{2015LRSP...12....6K}.     

In the photospheric umbrae, where plasma beta is close to 1 \citep{2017ApJ...837L..11C,sanjay2025flux}, acoustic waves may interact with the magnetic plasma and give rise to the MAWs. Alternate excitation mechanisms also exist for these waves, like broadband excitation \citep[e.g., see][]{2011ApJ...728...84B, 2015A&A...580A.107S} and magnetoconvection \citep{2017ApJ...836...18C}. The upward-propagating MAWs drive the umbral flashes \citep{1969SoPh....7..351B} observed in the chromosphere, which displays characteristic 3 minute periodicity \citep{2003A&A...403..277R}.

It is now widely accepted that these 3 minute oscillations are manifestations of slow MAWs guided along magnetic field lines \citep{2021ApJ...914...81K,2023MNRAS.525.4815R}. They can be observed directly from intensity oscillations because of their compressible nature.

Observations indicate that the 3 minute slow MAWs steepen while propagating upward and ultimately dissipate through the formation of shock \citep{2008A&A...479..213B,2017ApJ...844..129C,2018ApJ...854..127C}. This behavior explains findings from several studies \citep[e.g.,][]{2017ApJ...847....5K, 2024MNRAS.533.1166R, sanjay2025flux}, which show that the slow MAW amplitude increases up to a certain height and subsequently decreases. Motivated by these results, the present study focuses on investigating the steepening behavior or nonlinearity evolution of the 3 minute slow MAWs in the umbral atmosphere. 

The study of nonlinear acoustic waves in the solar atmosphere has initially been used to address the long-standing coronal heating problem, which was first recognized around the 1940s \citep{1939NW.....27..214G,1943ZA.....22...30E}. 
Pioneering studies like \citet{1946NW.....33..118B,1948ZA.....25..161B,1948ApJ...107....1S} proposed that the chaotic plasma motion in the convective zone acts as a source of upward-propagating acoustic waves, and due to density stratification, these waves steepen and become nonlinear as they travel and eventually form shocks.  The energy dissipation through these shocks was suggested as a mechanism for heating the corona.

More recently, \citet{2010ApJ...722..131F} applied a linear model to the phase difference and amplitude spectra of propagating slow MAWs in an isothermal stratified medium, incorporating radiative losses using Newton's law of cooling. Their results showed that nonlinear effects become increasingly significant in the upper chromosphere.   
\citet{2014ApJ...786..137T} observed a sawtooth pattern in the Doppler shift oscillations on the chromospheric and transition region spectral lines, suggesting upward propagation of slow magnetoacoustic shock waves.

Nonlinear effects in high-frequency acoustic waves were analytically described by \citet{naugolnykh1998nonlinear}. Although slow MAWs are commonly considered dispersionless, several authors \citep[e.g.,][]{1983SoPh...88..179E, 2017A&A...599A..15L, 2025MNRAS.538..797Z} have previously highlighted that these waves become dispersive in perpendicularly structured media. For an isothermal and gravitationally stratified atmosphere, \citet{2017A&A...599A..15L} derived both dispersive and nondispersive solutions of the nonlinear acoustic wave at the long and short wavelength limits and demonstrated the generation of 2nd harmonics analytically. They noted that the presence of the 2nd harmonic at twice the fundamental frequency is a robust signature of wave nonlinearity. Observational evidence for the presence of the 2nd harmonic was later provided by \citet{2017ApJ...844..129C}, using data from the Fast Imaging Solar Spectrograph \citep[FISS;][]{2013SoPh..288....1C} of the Goode Solar Telescope (GST) and by \citet{2018ApJ...854..127C} using the FISS and the Interface Region Imaging Spectrograph \citep[IRIS;][]{2014SoPh..289.2733D} as well. In addition, \citet{2009ApJ...692.1211c} reported the wave steepening in the chromosphere, based on the velocity profile of the He \textsc{i} 10830 \r{A} triplet, while \citet{2016ApJ...831...24K} observed similar behavior using temporal variation of line-of-sight velocity derived from the Mg II k line. However, all of these studies demonstrate the observational signature of wave nonlinearity at a single height and did not explore how this wave nonlinearity evolves with height across different atmospheric layers.

Building on this foundation, the present study observationally examines the nonlinearity of 3 minute slow MAWs across different atmospheric layers in the umbral atmosphere utilizing the data from the Helioseismic Magnetic Imager \citep[HMI;][]{2012SoPh..275..207S, 2012SoPh..275..229S} and Atmospheric Imaging Assembly  \citep[AIA;][]{2012SoPh..275...41B} onboard the Solar Dynamics Observatory  \citep[SDO;][]{2012SoPh..275....3P} for twenty different active regions.

In Section{\,}\ref{sec:data}, we provide a detailed description of the data utilized in this study; in Section{\,}\ref{sec:analysis} we describe the analysis methods and present our findings. Section{\,}\ref{sec:conclusion} concludes with a summary of key results and their implications.

\section{Data}\label{sec:data}

We analyzed multi-wavelength imaging data from HMI and AIA onboard SDO covering 20 different active regions observed between 2012 and 2016. This data set was initially used by \citet{2018ApJ...868..149K} and recently by \citet{2024ApJ...975..236S}, consisting of 1-hr long image sequences extracted using an IDL-based pipeline developed by Rob Rutten\footnote{https://robrutten.nl/rridl/sdolib/dircontent.html}. This routine co-aligns and derotates the images, corrects for instrumental effects, and upgrades level 1 to science-ready level 1.5.

Each dataset covered a 180\arcsec$\times$180\arcsec subfield centered on the sunspot umbra, with a consistent pixel scale of  0$\farcs$6. We analyze six channels $\-$- HMI continuum, AIA 1600, 1700, 304, 131, and 171 \r{A}, which sample the atmospheric layer from the photosphere to the lower corona. The cadence ranges from 45 to 12 seconds, depending on the channel. While the sunspot within our selected sample spans various Mount Wilson classifications, none exhibit significant flaring activity during the observational window, as verified by the Solar Monitor website\footnote{https://solarmonitor.org/}. However, eight sunspots show the presence of a light bridge and are indicated by asterisk symbols in our results presented in the following section. These regions are carefully excluded from our analysis to ensure that only the core umbral region is examined.

\section{Analysis and results}
\label{sec:analysis}

This study focuses on measuring the nonlinearity of slow MAWs with a 3 minute period propagating along vertical magnetic field lines in sunspot umbrae. The accurate identification of the umbral region in each sunspot is crucial to exclude the effects of the penumbral region in our wave analysis.

We determine the umbral region by utilizing the HMI continuum intensity images and applying a contrast-based segmentation method adopted from \citet{2017ApJ...847....5K}. This approach involves selecting a nearby quiet Sun region and computing its time-averaged median intensity. Some fraction of this median value is then used as an adaptive threshold for segmenting the umbra.

\begin{figure*}
    \includegraphics[width = 1\linewidth]{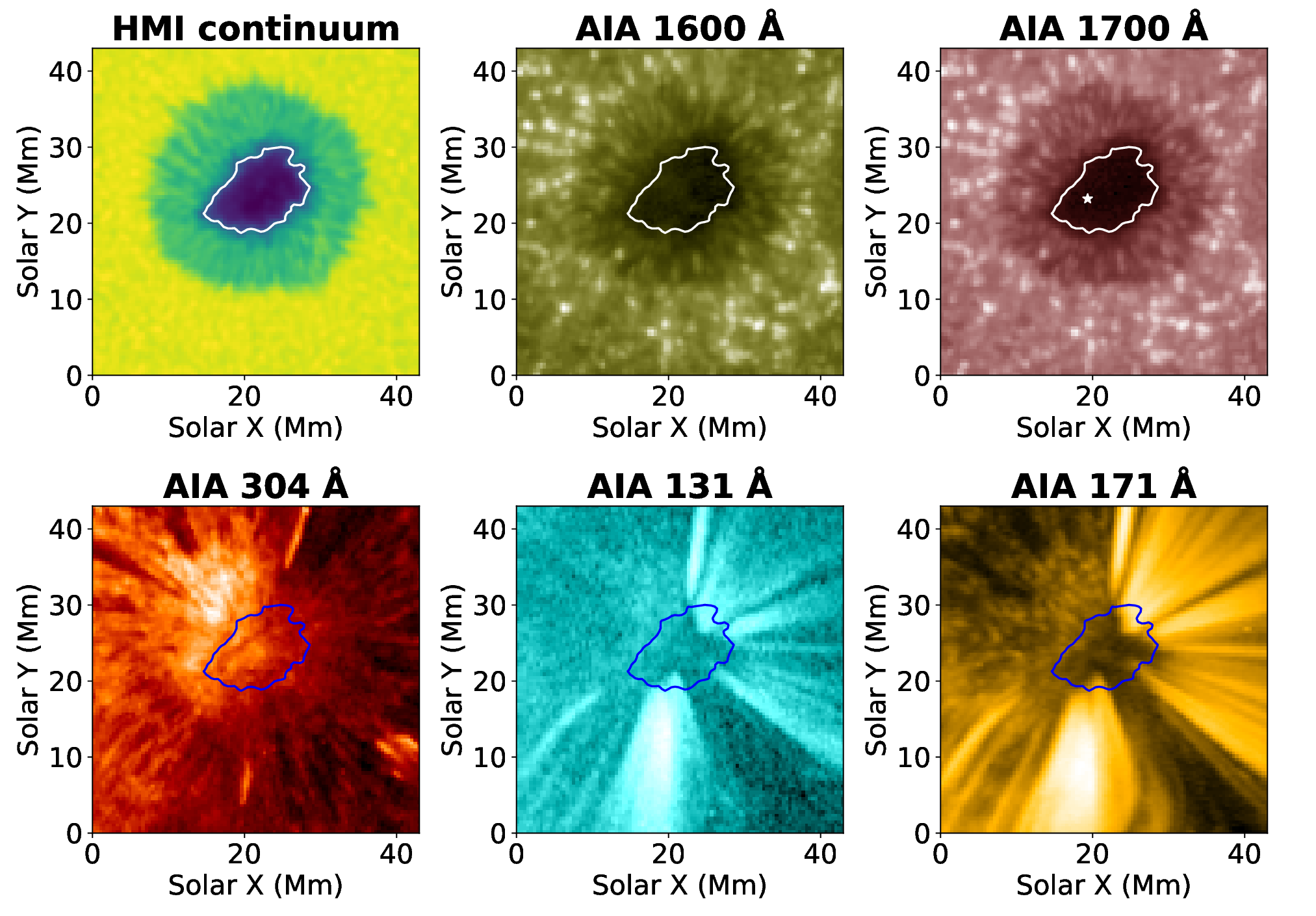}
    \caption{Sample images of a sunspot within AR 11591 captured on Oct. 19, 2012, across various wavelengths of SDO channels. The HMI continuum image depicts the time-averaged intensity, while the rest of the images represent the first frame of 1 hr duration data. The white and blue contours outline the umbra-penumbra boundary, and a white asterisk symbol in the AIA 1700 \r{A} image represents the test pixel for further processing.}
    \label{fig:umbra}
\end{figure*}
This intensity-based thresholding method naturally excludes the light bridge regions as well because they appear brighter than the surrounding umbra \citep{1997ApJ...484..900L}. As a result, the segmentation isolates only the darkest core, minimizing contamination from brighter structures. The threshold percentage is manually found for each active region to best match the observed visual boundary, accounting for variation in sunspot morphology and brightness. The chosen thresholds in our dataset range from 20\% to 55\% of the quiet Sun median value. The same umbral region was consistently chosen in all atmospheric layers because all our active regions are located not very far from the disk center, and as described in \citet{2024ApJ...975..236S}, the line-of-sight effect for this dataset is minimal. 

Fig.{}\ref{fig:umbra} represents the zoomed-in sample images of sunspots from AR 11591, observed on October 19, 2012. The umbral region is defined using a 40\% intensity threshold relative to the nearby quiet Sun median intensity. White and blue contours outline the umbra-penumbra boundary. 
\subsection{Wavelet Transform }\label{subsec:wavelet}

After isolating the umbral region from the surrounding penumbra and any light bridges, the long-term trend in the time series of each umbral pixel is removed by subtracting a 10 minute running average signal, retaining only periods shorter than 10 minutes. This step minimizes the effect of very low-frequency signals, ensuring that the analysis focuses on relevant oscillatory periods.
To investigate the temporal evolution and frequency characteristics of the observed oscillations, the time series data are transformed into the time–frequency domain using the wavelet transform technique described by \citet{1998BAMS...79...61T}. Specifically, we employed a complex Morlet wavelet and selected the smallest scale to be twice the data cadence. This choice corresponds to the shortest resolvable Fourier period according to the Nyquist criterion.

The wavelet power for each umbral pixel is normalized by its mean power. As an illustrative example, Fig. \ref{fig:wavelet} shows the normalized wavelet power spectrum obtained from the pixel location marked in the AIA 1700 \r{A} image of Fig.{}\ref{fig:umbra}.  This pixel was chosen near the umbra center in AIA 1700 \r{A} due to the presence of umbral flash signatures in the chromosphere, which indicate shock wave activity \citep{1997ESASP.404..189B,2014ApJ...795....9F} and wave nonlinearity. The wavelet spectrum reveals that oscillations with a period of $\approx$ 3 minutes are dominant for most of the time throughout the observation. This is consistent with the well-known presence of 3 minute oscillations in the chromosphere above sunspot umbrae. 

Furthermore, analysis of the wavelet power spectra reveals that the 3 minute oscillations do not persist uniformly throughout the observational period. Instead, they appear in the form of discrete wave packets, each lasting $\approx$ 10 to 20 minutes. This behavior is consistent with findings from earlier investigations \citep[e.g.,][]{2003ApJ...591..416C, 2015ApJ...812L..15K,2017ApJ...844..129C}. 

\subsection{Nonlinearity Index }\label{subsec:Nonlinearity}

 In order to measure the degree of nonlinearity of 3 minute slow MAWs, we make use of the relative strength of their 2nd harmonic \citep{2017ApJ...844..129C}. 
This captures how effectively the fundamental wave redistributes its energy into higher harmonics as a result of nonlinear interactions. As the amplitude of the fundamental oscillations increases, these nonlinear processes become significant, giving rise to a measurable rise in higher harmonics power.

We compute a time-dependent nonlinearity index (NI) for each umbral pixel across different atmospheric heights to capture this behavior quantitatively. This index is defined as

\begin{equation} 
NI = \sqrt{\frac{P_2}{P_1}}, 
\label{NI}
\end{equation}

where $P_1$ and $P_2$ represent the power of the fundamental mode and the 2nd harmonic, respectively. A key step in calculating this index is the precise identification of the period bands associated with these two modes.
\begin{figure*}
    \centering
    \includegraphics[width = \linewidth]{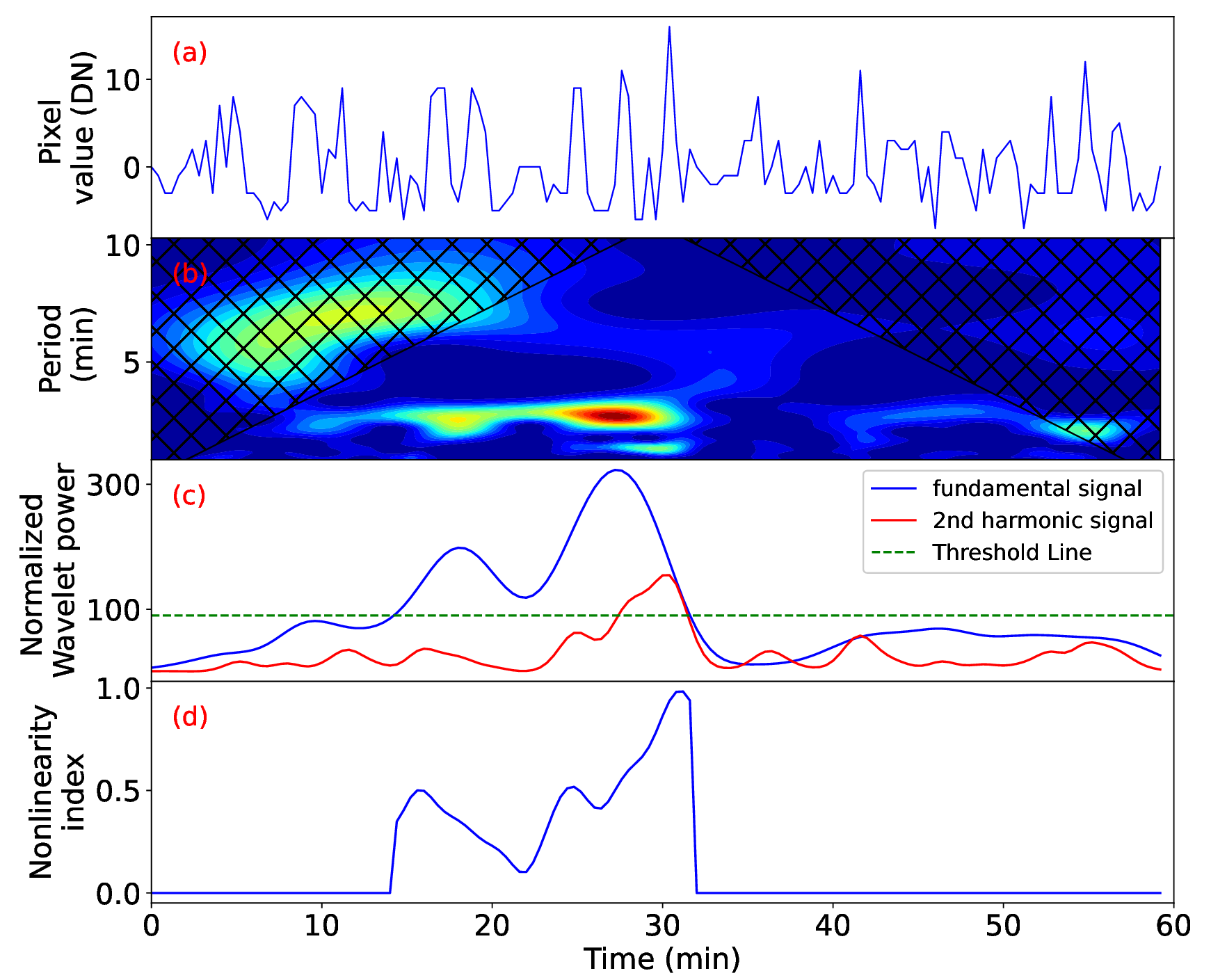}
    \caption{(a) It represents the detrended light curve of the pixel depicted in the AIA 1700  \r{A} image of Fig.{}\ref{fig:umbra} by an asterisk symbol. Its wavelet transform is shown in (b), along with the cone of influence. Blue and red lines depict the fundamental and 2nd harmonic power along with a dashed threshold green line in (c), which is defined in section \ref{subsec:Nonlinearity}. (d) represents its instantaneous NI.}
    \label{fig:wavelet}
\end{figure*}

\begin{figure*}
    \centering
    \includegraphics[width = \linewidth]{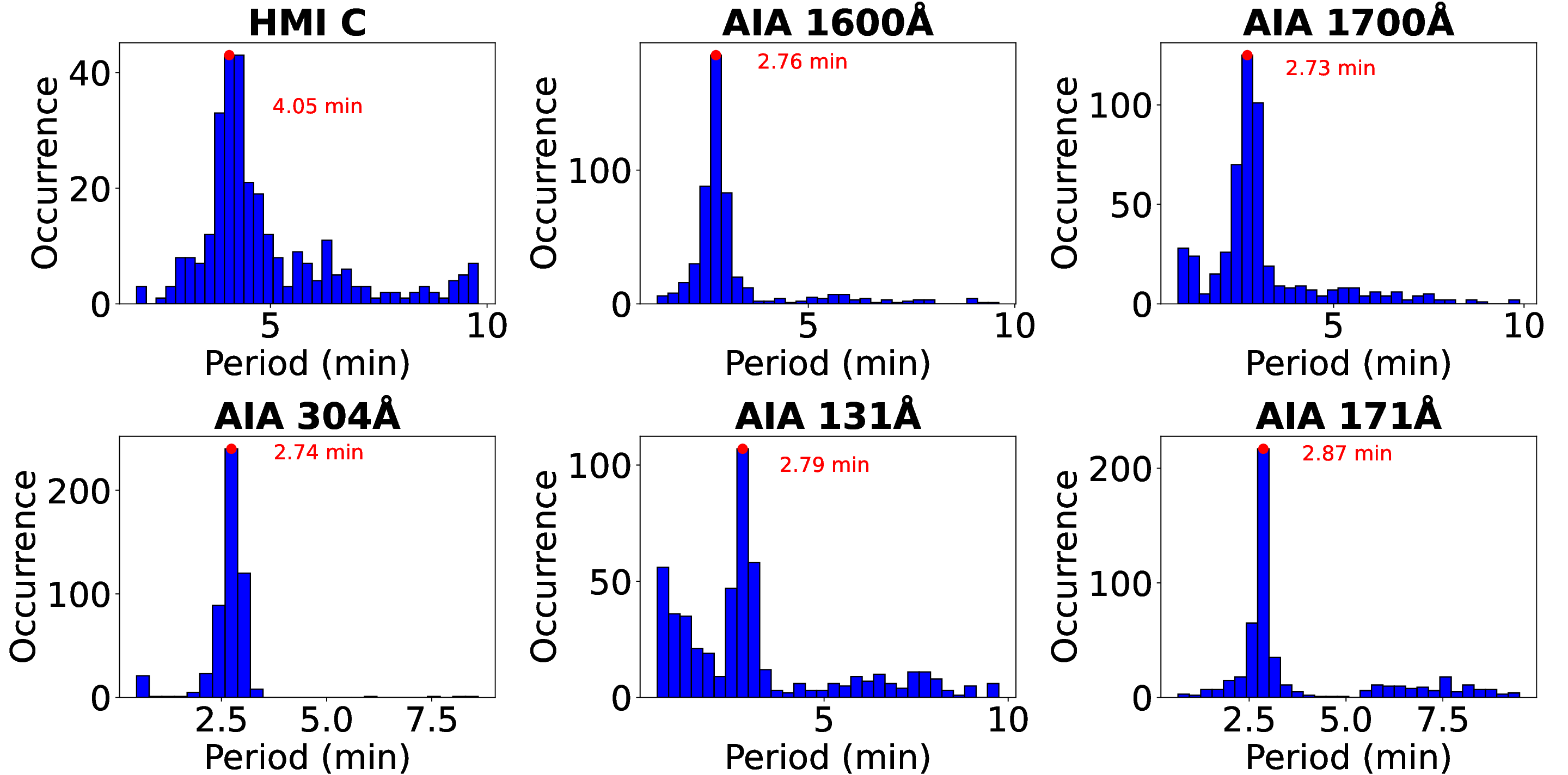}
    \caption{The dominant period for each channel for AR 11591 is displayed in the histogram, constructed using a fixed 20 second bin size. The center period of the bin with the highest occurrence, representing the most frequent dominant period, is also indicated in the figure.}
    \label{fig:period}
\end{figure*}

Since a single fixed period fails to adequately capture the complexity of acoustic waves, we employ a data-driven method to determine the period bands for the fundamental and 2nd harmonic signals. For each umbral pixel, the dominant period is identified as the frequency at which the wavelet power reaches its maximum. To obtain a statistically representative value for each wavelength channel, we construct a histogram of these dominant periods using a 20 second bin size per channel, applying the smallest scaling as twice the cadences in the wavelet transforms across all active regions. An example of such a histogram, corresponding to the AR 11591 active region, is shown in Fig.{}\ref{fig:period}. The center of the bin with the highest occurrence is adopted as the representative dominant period for that channel.

For AR 11591, the dominant periods identified for the HMI continuum, AIA 1600, 1700, 304, 131, and 171 \r{A} channels are 4.05, 2.76, 2.73, 2.74, 2.79, and 2.87 minutes, respectively. Consequently, the 2.80 minute period ($T_f$) is used as the central value for the fundamental period band for all the channels in this active region. This value varies from 2.7 to 2.9 minutes across all 20 active regions. 

The chosen range for the fundamental period band is from $\frac{T_f}{1.25}$ to $T_f \times 1.25$, which effectively captures the dominant oscillations. Similarly, the 2nd harmonic band spans the same relative range at half the fundamental central period. 
For example, the fundamental and its 2nd harmonic period bands for AR 11591 are [2.24, 3.50] and [1.12, 1.75] minutes, respectively, and the wavelet power within these bands is summed to estimate the fundamental and 2nd harmonic components for each umbral pixel. 
The same method is consistently applied across all observed active regions and channels to determine both signals.

The observational evidence of the presence of 2nd harmonic components ($\approx$ 1.5 minute oscillations) in addition to the primary 3 minute oscillations is shown in the wavelet power spectra in Fig.{}\ref{fig:wavelet}(c). These harmonics appear as distinct power enhancements around the 30 minute mark in the time series at frequencies approximately twice that of the fundamental mode and are prominent during periods of strong oscillatory activity in 3 minutes. Although traces of power are observed near 1 minute periods in AIA channels, higher-order harmonics (e.g., third, fourth) are not included in our analysis due to the limited temporal resolution of the observational data, which constrains the reliable detection and separation of such components.

These observations provide compelling evidence that the chromospheric plasma above sunspot umbrae, particularly at the formation heights corresponding to the AIA 1700 \r{A} passband, operates in a nonlinear regime.
These enhancements are not uniformly distributed but rather vary across different spatial locations and evolve over time, indicating that nonlinearity in the 3 minute oscillations is both spatially and temporally dependent.

The extracted power signals corresponding to the fundamental and 2nd harmonic components are illustrated for the test pixel in Fig.{}\ref{fig:wavelet}(d). Since a simple direct ratio of their amplitudes for NI estimation yields misleading results, particularly when both signals have low powers, we employed a minimum threshold. Specifically, the threshold for each pixel is defined as the time-averaged value of the fundamental signal, ensuring that the NI is calculated only when the slow MAW power is above this threshold. 

\begin{figure*}[ht]
    \centering
    \includegraphics[width = \linewidth]{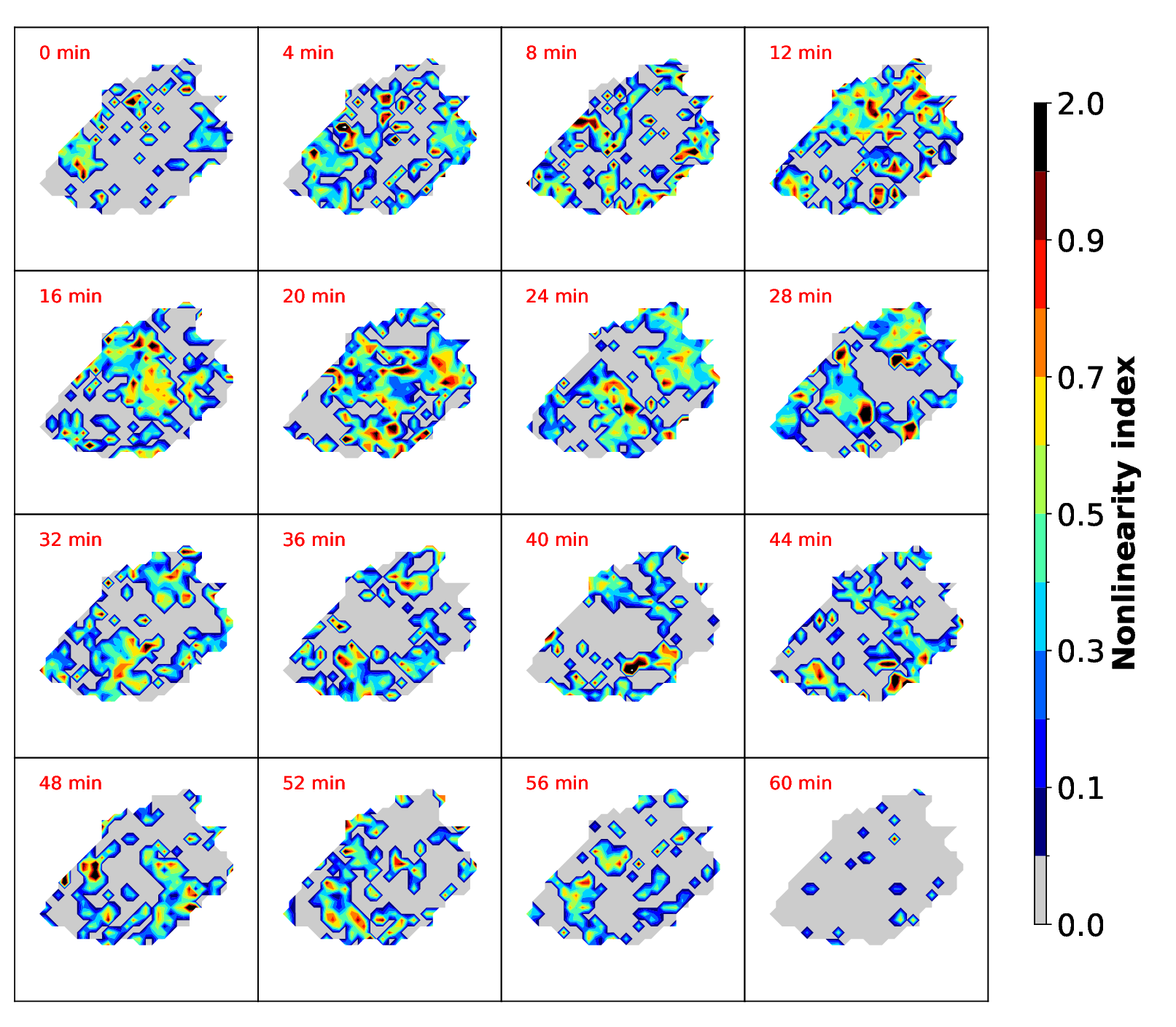}
    \caption{The spatial distribution of nonlinearity within a zoomed-in view of the umbra for the AIA 1700  \r{A} channel of AR 11591 over a fixed four minute time interval. The gray-colored pixels (NI $=$ 0) correspond to locations where the fundamental component is lower than their thresholds.  The rest of the color scale represents actual nonlinearity in the umbra, with most pixels exhibiting NI values in the range of [0.3, 0.6], and only a few pixels have NI $>$ 1, represented by black colors.}
    \label{fig:nl17}
\end{figure*}

\subsection{Nonlinearity variation with space and height}\label{nlheight}
The spatial distribution of nonlinearity within the umbra evolves dynamically over time. Fig.{}\ref{fig:nl17} shows the NI variation in the umbral region of AR 11591 for AIA 1700 \r{A}.  Between 16 and 24 minutes, nonlinear activity concentrates near the umbral center, while it becomes more irregular at other times. 

Such non-uniform and time-varying distributions may indicate the propagation of wave fronts with maximum power oscillations from the umbra center. These waves may interact between themselves and the originating background umbral flashes, as proposed by \citet{2018A&A...618A.123S}. In these flashes, we can assume that the nonlinearity of the oscillations will have maximum levels. In the inner part of the umbra boundary, we also observed the nonlinear effects, but they are associated with the enhancement of local umbral flashes, which are the footpoints of the magnetic field lines.  All these variations in space and time can inform us about a possible connection between power and nonlinear oscillations. However, these variations may also be influenced by the chosen threshold for NI estimation and potential edge effects from the wavelet analysis at the starting and ending times. 
 \begin{figure*}[ht]
    \centering
    \includegraphics[width = \linewidth]{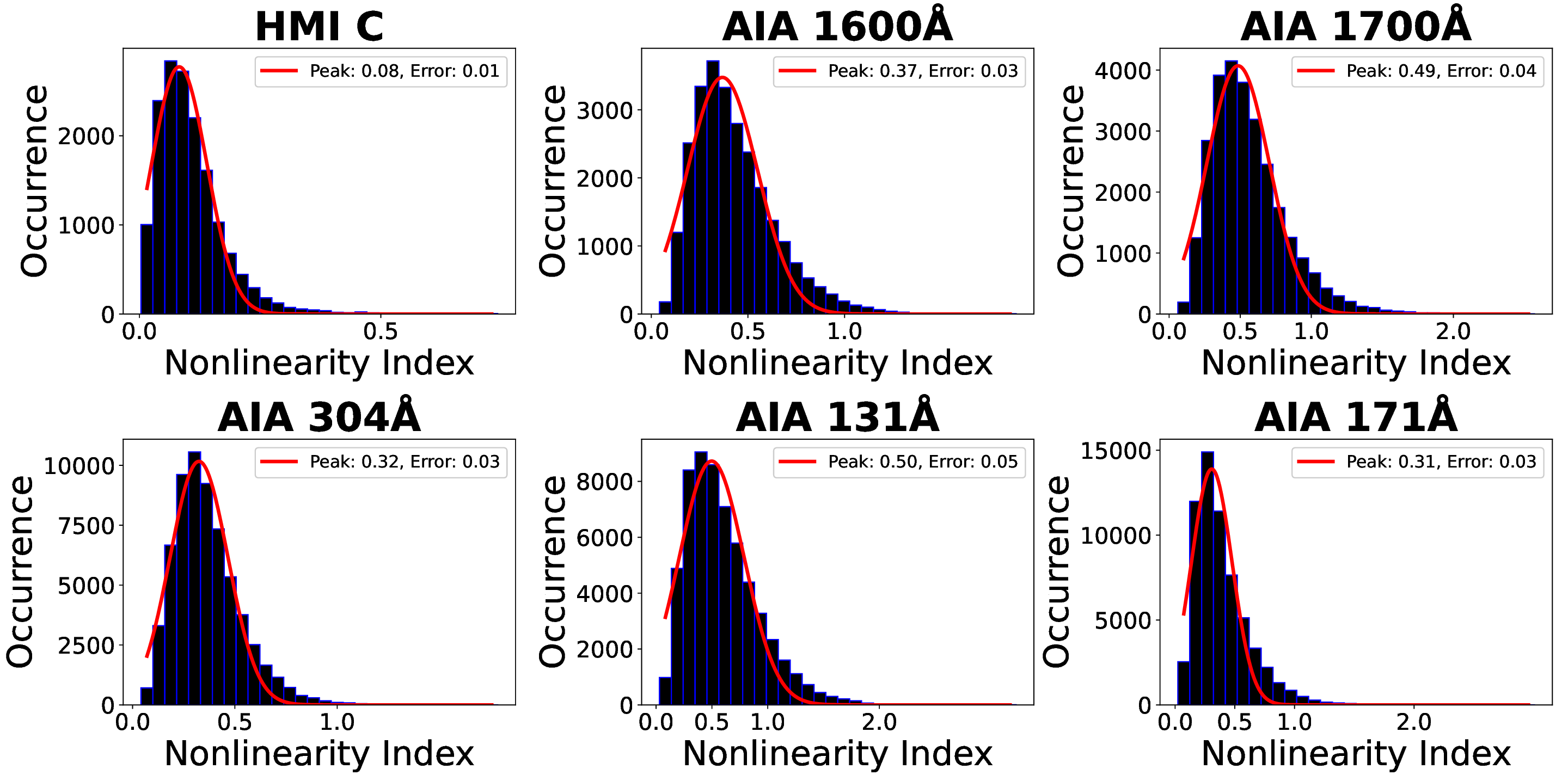}
    \caption{Histogram of all nonzero NI values obtained for all six channels of AR 11591. The red curves represent the Gaussian fit to the histogram data. The corresponding peak values and their standard errors are also mentioned in the plots.}
    \label{fig:nlhist}
\end{figure*}

We have also observed some occasional random spikes in the data, which contribute to artificially enhanced power at higher frequencies. Additionally, there are some cases where the threshold value for NI computation is low due to weak fundamental signals. Since the same fundamental period band is applied uniformly for all pixels, discrepancies arise for pixels where the actual dominant period deviates significantly from this reference value. It can lead to instances where the NI exceeds unity, indicating that the power in the 2nd harmonic surpasses that of the fundamental mode. In such cases, the signal likely identified as the 2nd harmonic is, in fact, the true fundamental mode for that particular pixel. 

The variation of NI with height is examined by constructing a histogram of nonzero NI values for each channel and fitting them with Gaussian curves to find the peak NI values and their standard errors, as shown in Fig.{}\ref{fig:nlhist} for AR 11591. These peak values are adopted as representative measurements of nonlinearity for each channel. It can be seen that the above-mentioned two effects are more prominent in the AIA 1700 \r{A} and AIA 131 \r{A} channels, where a larger number of events exhibit NI $>$ 1 compared to other channels. Similar findings, where 1.5 minute oscillation power is more than 3 minute oscillation power, were also reported by \citet{2018ApJ...854..127C}. However, in our analysis, the fraction of events where NI $>$ 1 is very small and statistically insignificant; thus, these outliers can be safely disregarded in the broader interpretation.

\begin{figure*}[ht]
    \centering
    \includegraphics[width = \linewidth]{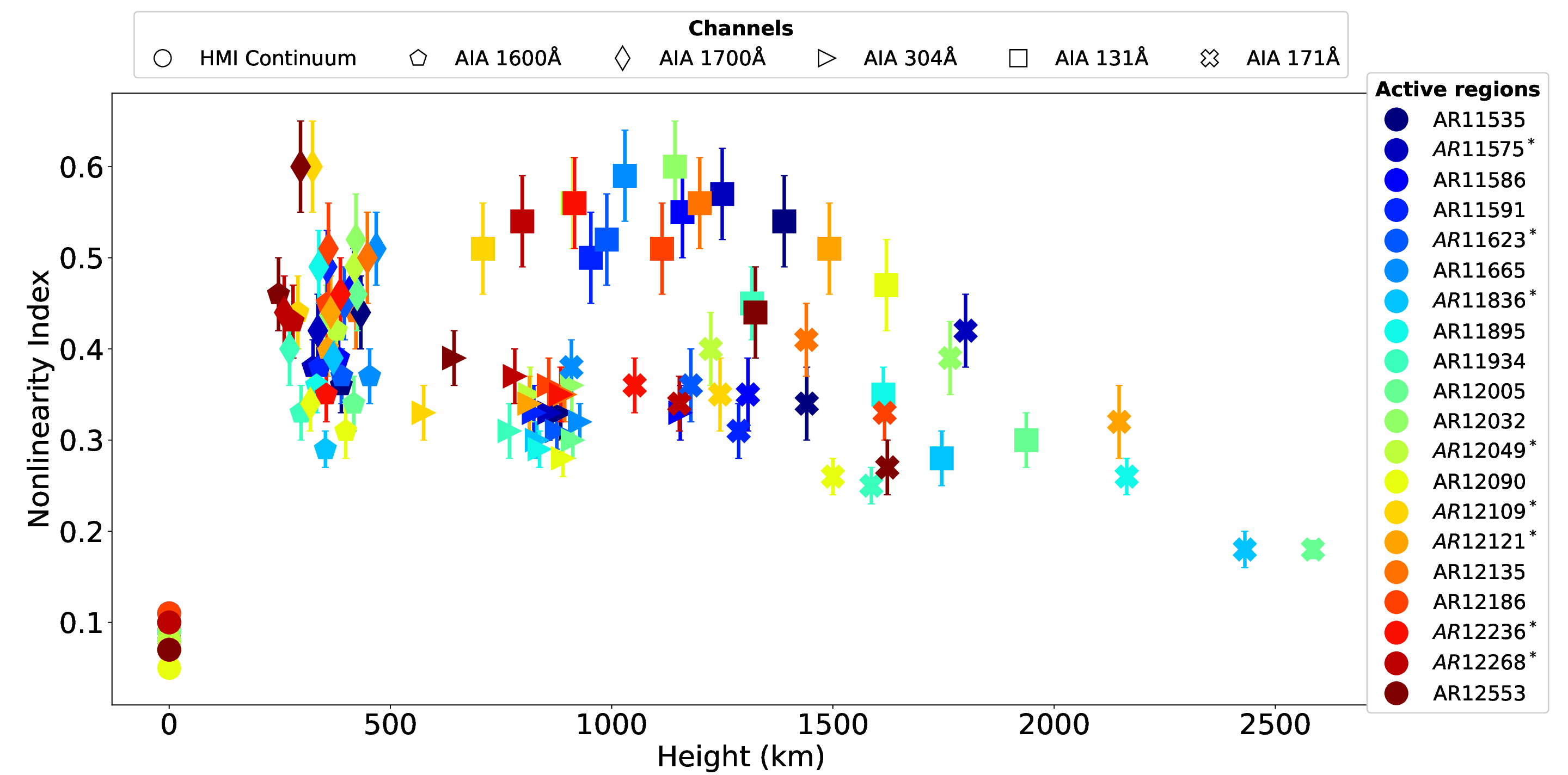}
    \caption{The variation of NI for six channels of HMI/AIA for all 20 analyzed active regions. Each color represents a different active region, while distinct symbols are used to denote different observational channels. The data points are arranged according to the estimated formation heights of each channel \citep{2024ApJ...975..236S}. Active regions that exhibit light bridges are marked with asterisks in the color legend.}
    \label{fig:nonlinearity}
\end{figure*}

The overall trend in NI across wavelengths reveals a distinct pattern: the nonlinearity increases up to the formation of AIA 1700 \r{A} channel, decreases at AIA 304 \r{A}, increases again at AIA 131 \r{A}, and subsequently decreases at the formation of AIA 171 \r{A}. The HMI continuum shows very low NI, representing that waves largely remain linear at the photospheric level. However, it must be noted that a portion of the chosen second harmonic period band lies very close to the Nyquist limit imposed by the 45 second cadence of the HMI continuum, which likely influences the estimation of the 2nd harmonic power and hence nonlinearity. Notably, this nonlinearity variation pattern is consistently observed across all 20 analyzed active regions, as shown in Fig.{}\ref{fig:nonlinearity}. In this figure, individual channels are represented by different symbols and active regions by distinct colors. All data points represented in this are according to the formation heights of the respective channels, as estimated by \citet{2024ApJ...975..236S}. In each case, there is a marked decrease in NI at the AIA 304 \r{A} level. One possible explanation is that the upward-propagating slow MAWs undergo shock dissipation near the height corresponding to AIA 304 \r{A} and consequently display a lower nonlinearity.

\section{Discussion and Conclusions}
\label{sec:conclusion}

Utilizing wavelet transform analysis, we estimated the nonlinearity of 3 minute slow MAWs by computing the powers in the fundamental and its 2nd harmonic across six channels of HMI and AIA for twenty distinct active regions. For this, we first determined the dominant period for each AIA channel within a given active region using the wavelet power spectrum. Based on these values, we then selected a single representative dominant period ($T_f$) to be used uniformly across all channels for that region. The NI was then calculated as the square root of the ratio of the total wavelet power in the 2nd harmonic band to that in the fundamental band above a threshold value. This index provides a measure of waveform distortion, with higher NI values indicating greater nonlinearity, typically associated with shock formation. It may be noted that we did not measure the nonlinearity directly from the steepening within the light curves because of the limited temporal cadence.

For the specific case of active region AR 11591, the wavelet analysis revealed dominant oscillation periods of [4.05, 2.76, 2.73, 2.74, 2.79, 2.87] minutes for the following channels, respectively: HMI continuum, AIA 1600, 1700, 304, 131, and 171 \r{A}. This result is consistent with the well-established frequency transition from 5 minutes in the photosphere to 3 minutes in higher layers, driven by the acoustic cutoff frequency \citep{horace1909theory, BellLeroy1977,2006ApJ...640.1153C} or by resonance cavity effects \citep{1982SoPh...79...19T,2024MNRAS.529..967S}. Based on this analysis, a dominant period of 2.80 minutes was selected as the representative fundamental mode for this active region.

The observed NI variation pattern is consistent across all active regions. It increases with height from the HMI continuum (photosphere) to the formation height of AIA 1700 \r{A} through the AIA 1600 \r{A} channel, followed by a noticeable decrease at AIA 304 \r{A}. This trend can be interpreted as slow MAWs propagating upward from the photosphere into regions of decreasing density; their amplitudes grow due to energy conservation in stratified media. This leads to progressive wave steepening and the eventual formation of shock fronts. Once the shocks dissipate their energy, the wave amplitude and thus NI begin to decrease at higher altitudes. 

Since, in this study, we estimated the nonlinearity using intensity oscillation, it can be linked to the results of  \citet{sanjay2025flux}, who examined the variation in intensity amplitude with the same dataset. They reported that the amplitude of 3 minute slow MAWs peaks in the height range of 700 $\--$ 900 km, then steadily decreases. After reaching 1800 km, the amplitude increases again before decreasing once more. It is reasonable to infer that the wave amplitude is likely to be maximum between the formation heights of AIA 1700 \r{A} and AIA 304 \r{A}. This suggests that the waves may have already undergone partial or full shock dissipation before reaching AIA 304 \r{A} formation. As a result, while the wave amplitude at AIA 304 \r{A} may still be higher than at AIA 1700 \r{A} due to the wave being in a decaying phase, the corresponding nonlinearity could appear reduced as a consequence of post-shock evolution.

According to the formation height estimated by \citet{2024ApJ...975..236S}, AIA 1700 \r{A} corresponds to altitudes between 260 km and 468 km with a median value of 368 km, while AIA 304 \r{A} forms between 575 km and 1155 km with a median value of 858 km for these active regions. This suggests that full shock development likely occurs between 368 and 858 km above the photosphere.

 \citet{2010ApJ...722..131F} observed that slow MAWs become nonlinear between the formation of Fe{} I ($\approx$ 560 km) and Ca{} II H (958 $\-$-1108  km) spectral lines. Theoretical work by \citet{2017ApJ...844..129C} proposes that 3 minute slow MAWs steepen into shock waves around 853 km and also mentions that signatures of nonlinearity may appear at lower altitudes.

Our observational results support this view, as nonlinearity is clearly detectable at the formation height of both AIA 1600 \r{A} and AIA 1700 \r{A} channels, suggesting the onset of shock formation begins well below the upper chromospheric heights. In addition, we observe a secondary rise in NI at the AIA 131 \r{A} channel, followed by a decrease in AIA 171 \r{A}. This behavior points to a 2nd phase of nonlinear wave evolution between AIA 131 \r{A} and AIA 171 \r{A}, and it can be linked to a 2nd phase of amplitude increase in the lower corona (AIA 171 \r{A} formation height) previously observed by \citet{sanjay2025flux}. This is possible due to the varying density structure in higher layers. 

The previously obtained data on the occurrence of oscillations of the second harmonic of the fundamental 3 minute mode in the wavelet spectra are based on spectroscopic observations of individual spectral lines of small magnetic formations, such as pores and flocculae without developed penumbrae. For images of large sunspots, spatial inhomogeneity of umbra and penumbra is observed in the form of narrowband sources in the 3 minute period band \citep{2024MNRAS.529..967S}

We propose an alternative explanation for the change in the nonlinearity of oscillations in the sunspots. Based on the results by \cite{Sych_2025}, the separate narrowband sources in the umbra as resonant cavities, where the oscillations are amplified at certain frequencies, were revealed. These sources are located at different distances from the center of the sunspot umbra, depending on the cutoff frequency of the wave, forming a complex volumetric structure of waveguides in the magnetic bundle. 

When the wave activity of the sunspot increases in the form of the appearance of 3 minute wave trains \citep{2014A&A...569A..72S}, there is an amplification of broadband waves passing through the found resonance cavities with splitting of the signal into separate multiple harmonics of the fundamental mode. During this process, frequency drifts of pulses in wavelet spectra are observed. The beginning and end of the wave trains coincide with the beginning and end of the frequency drift. The maximum power of the wave train coincides with the minimal value of the frequency drift \citep{Sych2012drift} and symmetry of pulses. This means that we observe changes in the shape of pulses, which in turn reflect the nonlinearity of oscillations at different times.

When the power of a broadband signal is amplified, it is assumed that all harmonics in the spectrum will experience simultaneous amplification. The wavelet spectrum will show the appearance of a series of signals with multiple
high-frequency harmonics during the maximum of the wave trains. The shape of their pulses will change over time, with a sequential change in nonlinearity in the form of different slopes of the pulses and their maximum symmetry at
the peak of activity. When mixed together, these harmonics will change the shape of broadband signals and lead to the evolution of their nonlinearity over time.

In the future, we plan to obtain the evolution of the distribution of nonlinearity over the sunspot and compare it with the location of resonant cavities at different heights. This will provide additional information on the spatial and temporal relationship between the sources of significant harmonics and waves nonlinearity.

We anticipate that future studies incorporating IRIS and other complementary satellite/ground-based observations, capable of bridging the current observational gaps, will provide improved constraints on the vertical structuring of the nonlinearity of slow MAWs. It is also worth noting that our method, which focuses exclusively on the power contained in the fundamental and 2nd harmonic components, may underestimate nonlinearity in scenarios where higher-order harmonics are significant but remain unresolved due to limited temporal resolution. Higher cadence and spectral coverage in future missions will be crucial for capturing the full complexity of nonlinear wave phenomena in the solar atmosphere.

\begin{acknowledgements}
YS gratefully acknowledges the financial support provided by UPES, Dehradun, India. SKP is grateful to SERB/ANRF for a startup research grant (No. SRG/2023/002623). The work of RS was financially supported by the Ministry of Science and Higher Education of the Russian Federation and the Chinese Academy of Sciences President’s International Fellowship Initiative, PIFI Group grant No. 2025PG0008. The authors are grateful to the SDO/AIA/HMI teams for operating the instruments and performing basic data reduction, and especially for their open data policy.
\end{acknowledgements}

\bibliography{nl}{}
\bibliographystyle{aasjournal}

\end{document}